# RevMine: An LLM-Assisted Tool for Code Review Mining and Analysis Across Git Platforms

Samah Kansab
*Software and IT Engineering Department*
*École de technologie supérieure (ÉTS)*
Montreal, Canada
samah.kansab.1@etsmtl.net

Francis Bordeleau
*Software and IT Engineering Department*
*École de technologie supérieure (ÉTS)*
Montreal, Canada
francis.bordeleau@etsmtl.ca

Ali Tizghadam
*Network Softwarisation and AI*
*TELUS*
Toronto, Canada
ali.tizghadam@telus.com

*Abstract*—Empirical research on code review processes is increasingly central to understanding software quality and collaboration. However, collecting and analyzing review data remains a time-consuming and technically intensive task. Most researchers follow similar workflows—writing ad hoc scripts to extract, filter, and analyze review data from platforms like GitHub and GitLab. This paper introduces RevMine, a conceptual tool that streamlines the entire code review mining pipeline using large language models (LLMs). RevMine guides users through authentication, endpoint discovery, and natural language–driven data collection, significantly reducing the need for manual scripting. After retrieving review data, it supports both quantitative and qualitative analysis based on user-defined filters or LLM-inferred patterns. This poster outlines the tool's architecture, use cases, and research potential. By lowering the barrier to entry, RevMine aims to democratize code review mining and enable a broader range of empirical software engineering studies.

## I. INTRODUCTION

Code reviews are a cornerstone of modern software development, enabling collaborative evaluation of changes, improving code quality, and facilitating knowledge sharing. Consequently, code review data has become a key subject of investigation in empirical software engineering. Researchers increasingly analyze pull requests and merge requests to study various development aspects, such as developer behavior, review practices, project dynamics, and opportunities for automating code review (see related work[1]).

Despite its value, the process of collecting and analyzing code review data remains technically demanding. Researchers often repeat similar workflows: manually authenticating to platforms like GitHub or GitLab, discovering and configuring relevant API endpoints/queries, writing scripts to extract and filter data, and applying custom logic to perform analysis. These tasks are not only time-consuming, but they also introduce inconsistencies and limit reproducibility across studies.

This paper presents RevMine, a conceptual tool that streamlines the code review mining process by leveraging large language models (LLMs) to bridge the gap between research intent and technical execution. Unlike existing solutions such as PyDriller or GH Archive, RevMine provides a no-code, interactive pipeline that unifies data collection, transformation, interpretation, and visualization through an intuitive dashboard. Users can define their analysis goals using natural language—for example, retrieving reviews from a specific timeframe or matching certain patterns—and RevMine automatically infers the necessary API calls and filters. Supporting both GitHub and GitLab, the tool facilitates cross-platform analysis and delivers both quantitative and qualitative insights.

In this poster, we present the motivation, architecture, and potential of RevMine. Our goal is to empower researchers with a flexible and intelligent assistant that streamlines review mining tasks and opens up opportunities for broader and more accessible empirical studies in software engineering.

## II. RELATED WORK

Code review is a critical quality assurance activity in modern software development. It is used to detect defects, improve code readability, and ensure compliance with design and style guidelines. Empirical evidence shows that reviewed code tends to be of higher quality and less likely to contain bugs [16, 2, 24]. RevMine builds on this foundation by enabling researchers to efficiently mine, filter, and analyze review data across platforms using LLM-guided orchestration and structured metrics.

Several studies have modeled the dynamics of the code review process, identifying factors that affect review duration and reviewer behavior. For example, predictive models for review completion and reviewer engagement have been proposed by Chouchen et al. and Maddila et al. [6, 15], while Hasan et al. [7] show that fast initial responses (particularly from humans rather than bots) correlate with shorter review lifetimes. Social and contextual factors also play a role: Bosu et al. [4] show that well-known developers receive faster feedback, while Jiang et al. [9] link the number of reviewers to review duration. Other studies focus on inefficiencies and bottlenecks, such as patch features that attract more attention [23], duplicated pull requests [14], and abandoned submissions [12]. Notably, Kansab et al. [11] demonstrate that mining review activity can expose deeper DevOps patterns, including effects of process changes and branching policies—a use case RevMine aims to directly support through customizable metric collection.

---
[1] https://figshare.com/s/c5b0e108520072210818

The usefulness and quality of review comments are also a focal point in the literature. Yang et al. [26] introduced EvaCRC, a framework for explainable comment quality assessment. Turzo et al. [25] conducted a large-scale study in the OpenDev Nova project, showing that linguistic clarity, politeness, and reviewer expertise contribute to perceived usefulness. Rahman et al. [19] developed RevHelper, a model predicting comment usefulness based on textual and contextual features, achieving 66

Other researchers have examined how different types of artifacts are treated during reviews. For example, Nejati et al. [17] show that build specification files are reviewed less often but draw more defect-related feedback. Spadini et al. [21] find that reviewers prioritize production over test code when both are changed, although test files still serve a critical verification function. Similarly, Thongtanunam et al. [22] report that defect-prone files often receive less reviewer attention, but raised issues often target evolvability, such as improving documentation rather than correcting functional logic. These findings inform RevMine's support for file-type-based filtering and classification during analysis.

In the industrial context, code review practices have been studied to understand scalability and tool support. Bosu et al. [5] conducted a multi-phase study at Microsoft, revealing how reviewer tenure and review scope affect comment usefulness. Hasan et al. [8] explore developers' perceptions of feedback and propose automated models for evaluating comment quality. Jureczko et al. [10] studied review effectiveness at Ocado Technology, comparing tool-assisted and over-the-shoulder reviews. They found that while automated tools produce higher volumes of comments, in-person reviews improve knowledge sharing. Alomar et al. [1] analyzed how developers at Xerox review refactoring changes, pointing to challenges in conveying intent and understanding structural impact. Kim et al. [13] investigated automated code review bots at Samsung Electronics, showing their effectiveness in detecting style issues and speeding up the review process. Likewise, systems like Google's Tricorder [20, 18] and Microsoft's internal code analysis tools [3] demonstrate that automation enhances review efficiency without replacing human judgment.

Together, these studies underscore the need for tooling that bridges manual and automated review processes, supports multi-dimensional analysis, and adapts to different artifact types and review contexts. RevMine addresses this gap by offering an LLM-guided, platform-agnostic tool that streamlines data collection and supports both quantitative and qualitative review analysis across a variety of research and industrial scenarios.

## III. Tool Architecture

RevMine is built as a modular system that enables researchers to extract and analyze code review data across platforms in a guided, semi-automated manner. As shown in Figure 1, its architecture consists of four main components: (1) Platform and Access, (2) Orchestrated Interaction, (3) Data Collection, and (4) Data Analysis Engine.

**1. Platform and Access.** The user begins by selecting GitHub or GitLab and providing access credentials (personal access token, project/repository ID, and optional metadata for self-hosted instances). RevMine then automatically verifies permissions, endpoint availability, and token validity to ensure smooth and error-free data collection.

**2. Orchestrated Interaction.** Once the platform is configured, users define their data collection intent through a flexible interaction layer that supports both natural language queries and manual configuration. For example, a user may write: *"Collect the commits of all the merge requests created in 2023 that include at least one reviewer comment."* RevMine forwards this prompt to an integrated LLM, which identifies relevant API endpoints (e.g., listing merge requests by creation date, retrieving associated commits, fetching comments), infers the appropriate filters (e.g. presence of comments), and determines which metrics should be extracted. It then generates a preliminary data collection plan and presents it to the user for validation or refinement before execution. We chose to integrate LLMs to allow users to express complex review mining goals in natural language. Unlike fixed-rule systems or custom scripts, LLMs enable exploratory and flexible data collection across a wide range of project structures.

A manual configuration mode remains available for users who prefer deterministic, reproducible setups. In this mode, users can select desired metrics through a structured graphical interface. These metrics are organized into intuitive categories (e.g., commits, comments, metadata), enabling users to include an entire group of related fields with a single selection. For example, choosing the `commits` category will automatically include metrics such as creation date, authored date, author identity, commit message, and file diffs, without requiring individual selection.

This interaction layer bridges the gap between high-level research questions and low-level API configurations. It empowers users—especially those unfamiliar with Git platform internals or REST APIs—to define and launch complex data workflows without writing any code.

**3. Data Collection.** Once the data collection plan is confirmed, RevMine initiates the retrieval process by issuing platform-specific API requests to extract structured code review data. This can includes metadata for each merge or pull request (e.g., title, description, creation date, merge status), the list of changed files, associated commits, and both inline and general review comments. The system is built to robustly handle pagination, rate limiting, and transient errors through automatic retries and detailed logging, ensuring reliable and traceable data extraction.

To support reproducibility and reduce redundant API usage, RevMine first stores all retrieved data in its raw form as a JSON file. This allows researchers to revisit or reprocess the data later without needing to re-run the full extraction pipeline—saving time and avoiding platform quota limitations.

Following raw data acquisition, RevMine constructs a structured dataset based on the specific metrics selected by the user (see the evolving list of metrics in[1]). At this stage, additional

filters—such as file extensions, time windows, or user-defined keywords—can be applied to refine the dataset. The resulting data is exported in CSV format and made accessible for further analysis or visualization.

**4. Data Analysis.** After data collection, RevMine offers an interactive analysis interface that supports both predefined metrics and user-driven exploration. Users can generate high-level summaries based on the collected metrics—such as the average number of comments per review, review duration, or basic metric trend charts—or request the LLM to generate custom analysis code (e.g., Python scripts that group reviews by week and visualize comment activity). The generated outputs can be displayed within the tool as part of an interactive dashboard and exported for reporting or publication.

This flexibility allows users to iteratively refine their research questions and adapt the analysis process without rewriting scripts. RevMine supports both quantitative insights (e.g., counts, durations, participation) and qualitative filtering based on review comment content, using keyword matching or LLM-inferred patterns.

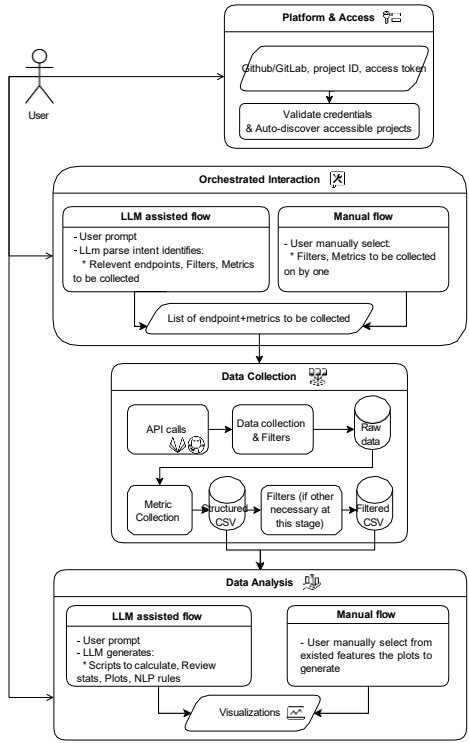

Fig. 1. Conceptual architecture of the RevMine tool

## IV. TESTS & VALIDATION

We are collaborating with two industrial partners—large software companies with active development teams—to test and validate the tool in real-world settings. Experts from these companies will interact with RevMine, provide structured feedback on usability, and evaluate its relevance for their internal analytics workflows.

The tool will be deployed within the partners' development environments for validation. In parallel, we are applying RevMine to open-source projects to compare its outputs with existing scripts and benchmarks, assessing consistency and reliability. We also plan to extend support to additional platforms such as Gerrit to broaden its applicability.

## V. CONCLUSION & PROGRESS

Overall, *RevMine* offers a complete, customizable, and user-friendly pipeline that eliminates the need for repetitive scripting while supporting high-quality and reproducible empirical research on code reviews across Git-based platforms.

To date, we have implemented core components of the tool, including the data collection engine, metric calculation routines, metric grouping logic, and initial testing of the LLM-based orchestration layer. The next steps involve evaluating and comparing LLM-based orchestration strategies, fine-tuning LLMs for context-aware code generation (e.g., generating plots or analysis scripts), and completing the user interface for interactive usage.